\documentclass{Odyssey2026}

\usepackage[T1]{fontenc}
\usepackage[utf8]{inputenc}
\usepackage{siunitx}
\usepackage{comment}
\usepackage{cite}
\interspeechcameraready 

\title{Evaluating voice anonymisation using similarity rank disclosure}

\author[affiliation={1,3}]{Shilpa}{Chandra}
\author[affiliation={1,4}]{Matteo}{Pettenò}
\author[affiliation={1}]{Nicholas}{Evans}
\author[affiliation={1}]{Michele}{Panariello}
\author[affiliation={1}]{Massimiliano}{Todisco}
\newlineauthors
\author[affiliation={2}]{Tom} {B{\"a}ckstr{\"o}m} 
\author[affiliation={3}]{Dorothea}{Kolossa}
\author[affiliation={4}]{Rainer}{Martin}
\author[affiliation={5,6}]{Themos}{Stafylakis}
\author[affiliation={7}]{Nicolas}{Gengembre}

\affiliation{}{EURECOM}{France}
\affiliation{}{Aalto University}{Finland}
\affiliation{}{Technische Universität Berlin}{Germany}
\affbreak
\affiliation{}{Ruhr-Universität Bochum}{Germany}
\affiliation{}{Omilia}{Greece}
\affiliation{}{Athens University of Economics and Business}{Greece}
\affiliation{}{Orange}{France}

\email{\{firstname.lastname\}@eurecom.fr, tom.backstrom@aalto.fi, dorothea.kolossa@tu-berlin.de, rainer.martin@rub.de, tstafylakis@omilia.com, nicolas.gengembre@orange.com}

\usepackage{comment}
\usepackage{soul}
\usepackage{float}
\usepackage{subcaption}  
\usepackage{xcolor}

\begin{document}

\maketitle

\def\thefootnote{\fnsymbol{footnote}}
\footnotetext[1]{This work was performed while the second author was a PhD candidate at EURECOM and Ruhr-Universität Bochum.}
\def\thefootnote{\arabic{footnote}}

\begin{abstract}

    The evaluation of voice anonymisation remains challenging. Current practice relies on automatic speaker verification metrics such as the equal error rate (EER). Performance estimates dependent on the classifier and operating point provide an incomplete or even misleading characterisation of privacy risk. We investigate the use of similarity rank disclosure (SRD), an information-theoretic metric, which operates on feature representations rather than classifier decisions, providing a threshold-independent assessment of privacy and analysis of both average and worst-case disclosure. We report its application to speaker embeddings, fundamental frequency, and phone embeddings using 2024 VoicePrivacy Challenge systems. The SRD reveals privacy leaks and system-specific weaknesses missed by EER-based evaluation. Findings highlight the merit of representation-level metrics and demonstrate the potential of SRD as a flexible and interpretable tool for the evaluation of voice anonymisation.
\end{abstract}

\section{Introduction}

Smart devices and cloud services are nowadays constantly capturing and processing speech data~\cite{wu2021understandingtradeoffs, backstrom2025privacyspeechtechnology}.
Beyond voice identity, speech recordings can reveal sensitive attributes such as the speaker's age, gender, emotional state, etc.~\cite{nautsch2019preservingprivacyspeaker,backstrom2025privacyspeechtechnology}.
The pervasive capture and inherent sensitivity of speech data, coupled with evolving privacy regulation~\cite{nautsch2019gdprspeechdata} have stimulated growing research interest in privacy for smart speech technology. 
 
The VoicePrivacy Challenge (VPC)~\cite{tomashenko2020introducing}, was founded in 2020 to foster the development of voice anonymisation solutions. 
Anonymisation is used to substitute the voice in a speech recording with that of another speaker or a pseudo-voice, preventing it from being linked to the original identity.
Depending on the specific application, however, other utility-related attributes, such as the linguistic and paralinguistic content, should be preserved.

There is hence an inherent trade-off between privacy and utility. 
Perfect anonymisation can be achieved simply by removing the speech content entirely. 
This trivial solution, however, has no practical value; the result is functionally useless. 
While the evaluation of voice anonymisation systems hence requires careful assessment of the balance between privacy protection and utility preservation~\cite{tomashenko2024voiceprivacy2024challenge}, our focus in this paper is exclusively upon evaluation of the former.

Privacy protection is usually evaluated  using automatic speaker verification (ASV) and related metrics. 
For the VPC, the \textit{Equal Error Rate (EER)} serves as the primary privacy metric. The EER and other related metrics, however, depend fundamentally on not just anonymisation performance, but also the ASV model. Estimates of privacy can then vary with the choice of ASV system, the features or embeddings employed, the training data with which it is trained, the distance measure, the operating point, etc. As a result, estimates of privacy can then be as much a function of the ASV system used for evaluation as of voice anonymisation performance.

We have investigated the use of a recently proposed privacy metric, the \textit{Similarity Rank Disclosure (SRD)}~\cite{backstrom2025privacydisclosuresimilarity}, for the evaluation of voice anonymisation solutions. 
The SRD is not dependent upon the binary decisions of an ASV system or similar classifiers and
can instead be applied directly to \emph{any} features or embeddings likely to capture speaker-related information.
The SRD can also provide estimates of average and worst-case privacy disclosure in information bits and could be applied readily to the evaluation of privacy leakage from the disclosure of other soft attributes beyond those immediately related to identity.

To demonstrate its flexibility, we demonstrate use of the SRD to estimate anonymisation performance and privacy disclosure from speaker embeddings, fundamental frequency and phone distributions~\cite{decheveigne2002yinfundamentalfrequency,niekerk2020vectorquantizedneuralnetworks,desplanques2020ecapatdnnemphasizedchannel,bakari2025influencenontimbralcues}. Using 2024 VPC protocols and data, we furthermore show that the SRD can reveal anonymisation weaknesses that are missed by evaluations performed using ASV systems and EER estimates.

The remainder of the paper is structured as follows. 
In Section~\ref{section:background}, we describe the existing metrics in voice privacy and current approaches to the evaluation of voice anonymisation systems.
In Section~\ref{section:SRD}, we describe our use of the SRD for evaluation. 
Our experimental setup is described in Section~\ref{section:Exp Setup}.
We present results and discussion in Sections~\ref{section:results} and~\ref{sec:discussion}, followed by conclusions in Section~\ref{section:conclusion}.

\section{Related work}
\label{section:background}
The strength of any approach to privacy protection is usually estimated empirically according to a particular threat model \cite{rahman2024scenario} and simulated attacks launched to defeat the protection. 
The strength is then quantified according to some objective metric that indicates the attack success rate. 
In the case of voice anonymisation, the metric reflects the extent to which the original speaker of an anonymised utterance can be re-identified.
In practice, re-identification is formulated as a binary ASV detection task.

A number of objective metrics have been proposed and applied to the evaluation of voice anonymisation systems~\cite{maouche2020comparativestudyspeech}. 
The EER was adopted as the primary privacy metric since the first VPC held in 2020~\cite{tomashenko2022voiceprivacy2020challenge}. 
EERs might not, however, reflect the realistic cost model of a privacy attacker who might instead prioritise a lower rate of missed detections while tolerating a higher rate of false alarms. Moreover, it can be shown that, under calibrated scores, the EER corresponds to the upper bound of the total error rate~\cite{brummer21_interspeech,Kinnunen2023-tEER} and the worst possible decision policy of a privacy adversary~\cite{nautsch2020privacyzebrazero}.  Use of the EER can hence result in exaggerated estimates of privacy protection.

The \textit{log-likelihood-ratio cost function}~\cite{brummer2006applicationindependentevaluationspeaker} offers one solution to overcome this limitation. 
Derived from the total log-likelihood-ratio cost ($C_{\text{llr}}$), it isolates discrimination capability via optimal calibration, often using the pool adjacent violators algorithm. $C_{\text{llr}}^{\text{min}}$ provides a scalar summary of the receiver operating characteristic (ROC) convex hull, offering a robust, threshold-independent measure that nonetheless still correlates well with the EER in the case of voice anonymisation.

The \textit{Zero Evidence Biometric Recognition Assessment (ZEBRA)} framework~\cite{nautsch2020privacyzebrazero} quantifies privacy disclosure based upon the empirical cross-entropy (ECE). The expected privacy disclosure $D_{ECE}$ is an estimate of the average information disclosed to an adversary in bits, while the worst-case privacy disclosure quantifies the maximum evidence revealed for any individual.

While the $C_{\text{llr}}$ and ZEBRA focus on information-theoretic aspects of privacy, the Linkability and Singling Out metrics introduced in~\cite{barrero2018generalframeworkevaluate, vauquier2025legallyvalidatedevaluation} provide a complementary, legally grounded perspective. 
Drawing on the GDPR and related privacy regulation, they capture residual re-identification risks that conventional measures may overlook. Linkability quantifies the probability that an anonymised sample can be correctly linked to the original speaker. Singling Out reflects the likelihood that an individual can be uniquely isolated from anonymised data, addressing risk even when identity remains unknown.

As we shall see, the SRD combines the merits of existing metrics into a single, flexible framework for the evaluation of voice anonymisation solutions and provides an interpretable understanding of how much personal identifying information (PII) remains in different representations of protected speech data.

\section{Similarity Rank Disclosure}
\label{section:SRD}

The SRD provides a framework to measure PII contained within speech utterances~\cite{backstrom2025privacydisclosuresimilarity}.
Information is quantified in bits, hence enabling comparisons between the disclosure of information contained in different speech characteristics. Operating directly upon speech features instead of classifier decisions, the SRD provides a classifier threshold-independent and entropy-grounded measure of privacy disclosure. With a full definition of the SRD available in the original work~\cite{backstrom2025privacydisclosuresimilarity}, we provide in the following only an overview of the SRD and necessary prerequisites relevant to its application as a metric for the evaluation of voice anonymisation solutions.

\subsection{Similarity evaluation and empirical ranking}
\label{sec:srdSetup}

Computation of the SRD involves comparisons between feature representations extracted from input utterances $x$ to a set of representations extracted from a database of $N$ reference utterances $y$.  The database contains one reference which matches the voice identity in each input $x$.  We then proceed as follows.

\begin{enumerate}
 
    \item \textbf{Ranking -} For every input $x$, we evaluate the similarity to each reference $y$. The set of resulting similarities is rank-ordered from the most similar (rank 1) to least similar (rank $N$).  For each $x$, we then determine the rank $k$ of the matching reference. 

    \item \textbf{Distribution generation -} A histogram of matching ranks $k$ is derived from a set of inputs $x$.  The histogram is normalized as an \textit{empirical probability distribution} $\tilde{p}_k$ of matching speaker rank.
\end{enumerate}

\begin{figure}[!t]
\centering
\includegraphics[width=0.55\linewidth]{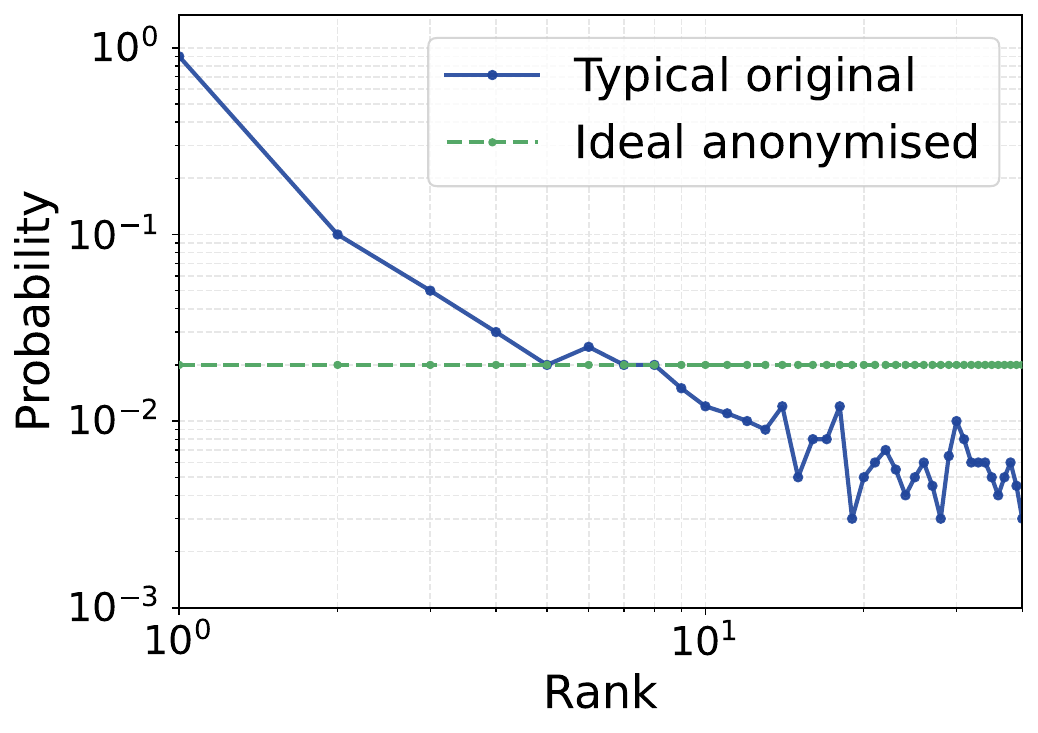}
\caption{Typical and ideal rank histogram distributions for original and anonymised data. }
\label{fig:ideal_rank_distribution}
\end{figure}

Example empirical probability distributions are shown in Figure~\ref{fig:ideal_rank_distribution} for a database containing $N=40$ references.
The blue profile illustrates a typical distribution for original, unprotected data.  
The largely-negative slope indicates a greater probability of matching speakers being in lower than in higher rank positions. 
The voice in inputs $x$ are readily identified and matched to the corresponding reference in the database.
Lower probabilities for higher ranks imply little confusion between speakers in the input and other, non-matching references.

The green profile illustrates the ideal distribution expected for perfectly protected data.  
This distribution should be uniform ($\tilde{p}_k=0.025$ for $N=40$); the voices in the inputs $x$ are as likely to correspond to the matching reference as to \emph{any} other speaker in the database.
In practice, protection is expected to be imperfect thus we expect distributions between those for unprotected and perfectly protected data.

\subsection{Statistical modelling }
\label{sec:beta}

With plentiful data, the normalised histogram $\tilde{p}_k$ can be used directly for further analysis. 
If, however, data is sparse, then
further analysis can be performed using statistical, parametric distributions estimated from empirical distributions.
The original work~\cite{backstrom2025privacydisclosuresimilarity} suggests use of the beta-binomial distribution.

\begin{enumerate}
    \item \textbf{Beta-binomial fit -} The fitting of a \textit{beta-binomial distribution} to the empirical distribution $\tilde{p}_k$ supports the estimation of smoothed probabilities $\gamma_j$ that input $x$ corresponds to the $j^{th}$ best match.
    
    \item \textbf{Parameter estimation -} 
    Due to a high penalty for distribution deviations at the crucial rank-1 position, for which disclosure is highest, model parameters $\alpha$ and $\beta$ are optimised using the \textit{constrained log-likelihood} loss function. 
\end{enumerate}

\subsection{Disclosure quantification}
\label{sec:sdr_metrics}

The rank order disclosure $\epsilon_j$ quantifies the reduction in uncertainty (entropy) gained from observing the matching rank $j$.

\begin{enumerate}
    \item \textbf{Rank order disclosure -} Before any observations, assuming all $N$ identities are equally likely as in the case of a uniform distribution, then the prior information needed to encode identity is $\log_2 N$ bits. After observing the rank $j$ of the matching reference, the posterior probability given by $\gamma_j$ can be estimated using either $\tilde{p}_j$ or a beta-binomial approximation. The information required is then reduced to $-\log_2 \gamma_j$ bits.
The \textit{rank order disclosure} $\epsilon_j$ given rank $j$ is then expressed in bits as:
\begin{equation}\
\epsilon_j := - \log_2 \gamma_j - \log_2 N
\label{eq:rank_disclosure}
\end{equation}
The subtraction in (~\ref{eq:rank_disclosure}), rather than addition, results in an expression of disclosure rather than of entropy or uncertainty.
    
    \item \textbf{Statistical summary -} Further insights into privacy disclosure can be gained from rank order disclosure distribution statistics:

    \begin{itemize}
        \item \textbf{Mean Disclosure (MeanD) -} The average disclosure across all observations.
        \item \textbf{Maximum Disclosure (MaxD) -} The single, worst-case disclosure observed.
        \item \textbf{Identification Rate (IdR) -} The probability that the matching reference is ranked first ($\tilde{p}_1$).
        \item \textbf{Rank Spread -} The proportion of ranks whose probability exceeds the level of pure chance ($1/N$).
    \end{itemize}

\end{enumerate}

Use of the SRD provides a characterisation of privacy disclosure in bits.
By focusing directly on feature representations, the SRD can also be used to compare privacy disclosure across any suitable feature representation, e.g., the fundamental frequency, phone embeddings, or speaker embeddings.
By decoupling evaluation from classifier decisions, estimates of privacy disclosure better expose and quantify risks arising from information contained in the representation itself. We use the SRD to evaluate the privacy leakage in voice anonymisation.

\begin{figure*}[!t]
  \centering
  \includegraphics[width=0.9\textwidth]{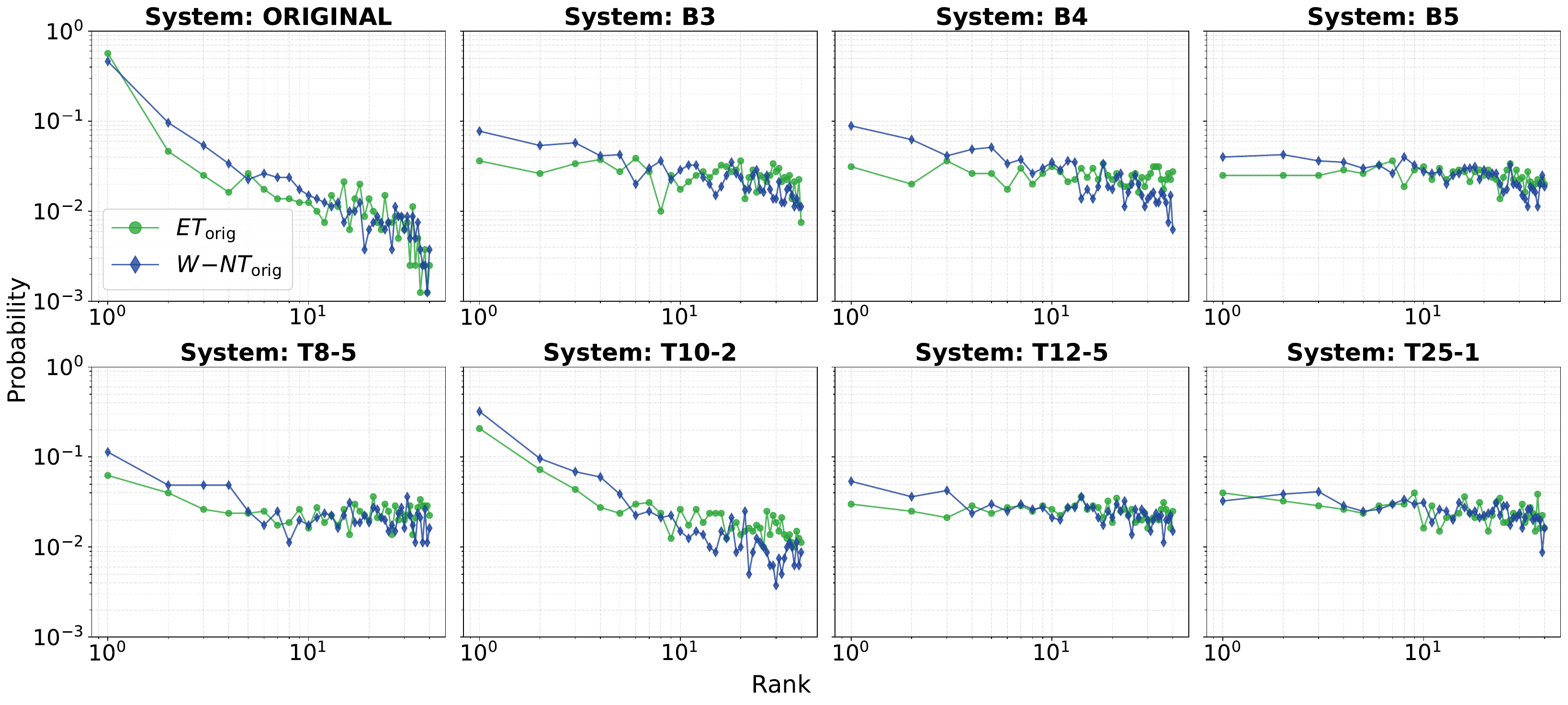}
  \caption{Rank histograms for the ECAPA-TDNN embeddings $ET_\text{orig}$ and non-timbral embeddings ${W\text{-}NT}_\text{orig}$, for original speech (top left plot only) and anonymised speech (all others).}
  \label{fig:ET_W-NT_org}
\end{figure*}

\section{Experimental Setup}
\label{section:Exp Setup}
While the original work~\cite{backstrom2025privacydisclosuresimilarity} reports a study of privacy disclosure for original, unprotected speech data, we have applied the SRD to the study and comparison of privacy disclosure for speech data treated with different approaches to voice anonymisation~\cite{tomashenko2024voiceprivacy2024challenge}.
In the following section, we describe the data used and a set of four feature representations.

\subsection{Anonymisation systems, data  and protocols}
\label{subsection:protocols}
We evaluate both baseline and top-performing 2024 VPC systems~\cite{tomashenko2024voiceprivacy2024challenge} also used for the 2024 VoicePrivacy Attacker Challenge~\cite{tomashenko2024firstvoiceprivacyattacker}. 
These include baselines B3, B4, and B5, and participant systems T8-5, T10-2, T12-5, and T25-1. We report results derived using the 2024 VPC evaluation set only.

VPC protocols define sets of enrolment and trial utterances, both of which may be anonymised.
Privacy protection is then estimated using an ASV system and sets of target and non-target trials, with each involving a comparison between one enrolment utterance and one trial utterance.
Using the SRD, instead of using ASV detection scores, we estimate privacy protection using similarity ranking.
The 2024 VPC protocols provide for only 29 speakers that are common to both enrolment and trial sets.
To provide for a larger number, we pool the sets of all enrolment and trial utterances (\textit{libri\_test\_enrolls}, \textit{libri\_test\_trials\_f}, and \textit{libri\_test\_trials\_m}) and construct custom, utterance-disjoint input and reference sets (see Section~\ref{sec:srdSetup}) following the procedures in~\cite[Sec.~IV]{backstrom2025privacydisclosuresimilarity}. 
This process yields 40 speakers common to both input and reference sets while also providing a greater number of observations per speaker.

\subsection{Feature representations}
\label{section: features}
We demonstrate the flexibility of the SRD with experiments using a selection of different feature representations known to capture speaker-dependent characteristics.
The selection is inspired by the features used in~\cite{backstrom2025privacydisclosuresimilarity}, namely the fundamental frequency, phone, and speaker representations.
For the latter, we use representations extracted using the same ECAPA-TDNN model~\cite{desplanques2020ecapatdnnemphasizedchannel} used in~\cite{backstrom2025privacydisclosuresimilarity} as well as for VPC evaluations \cite{tomashenko2022voiceprivacy2020challenge,tomashenko2022voiceprivacy2022challenge,tomashenko2024voiceprivacy2024challenge}. In addition, we make use of a WavLM model~\cite{Chen2022-wavlm} trained to use non-timbral cues~\cite{bakari2025influencenontimbralcues}.

It is well known that more reliable interpretations of protection can be derived using stronger attacks and embeddings extracted using models trained with similarly anonymised data. 
This approach, referred to as the \textit{semi-informed} attack model, is now standard practice for VPC evaluations~\cite{tomashenko2022voiceprivacy2020challenge, tomashenko2022voiceprivacy2022challenge,tomashenko2024voiceprivacy2024challenge}. 
We also use speaker embedding models trained this way.

We make no use of linguistic embeddings as in~\cite{backstrom2025privacydisclosuresimilarity}.
While also a source of speaker information, the VPC concerns strictly \emph{voice} characteristics rather than spoken content as sources of PII. 
None of the baseline systems, nor participant systems, manipulate linguistic information.\footnote{It is acknowledged that this might not be the case for baseline B3 and other ASR-TTS-based anonymisation systems.}
Furthermore, the LibriSpeech dataset consists of audiobook recordings sourced from the LibriVox project~\cite{panayotov2015librispeechasrcorpus}. 
It is hence a reasonable assumption that linguistic content reflects more the linguistic preferences of the book authors rather than those of the narrators/speakers.
Consequently, linguistic information is less likely to be a source of privacy disclosure than the other features used in our work.  

In the following, we describe the extraction of each feature representation used in our work and any differences to the implementations used in~\cite{backstrom2025privacydisclosuresimilarity}.

\subsubsection{ECAPA-TDNN (ET)}

We use the SpeechBrain~\cite{ravanelli2021speechbraingeneralpurposespeech} 
ECAPA-TDNN model~\cite{desplanques2020ecapatdnnemphasizedchannel} to extract speaker embeddings and
train two models.
The first, denoted $ET_\text{orig}$ is trained using original speech. 
The second, denoted $ET_\text{anon}$, is trained under the \textit{semi-informed} attack setting using anonymised speech, as described in Section~\ref{section:results}. 
The empirical probability distribution is derived from the cosine distance between embeddings extracted from input and reference pairs.

\begin{figure*}[!t]
  \centering
  \includegraphics[width=0.9\textwidth]{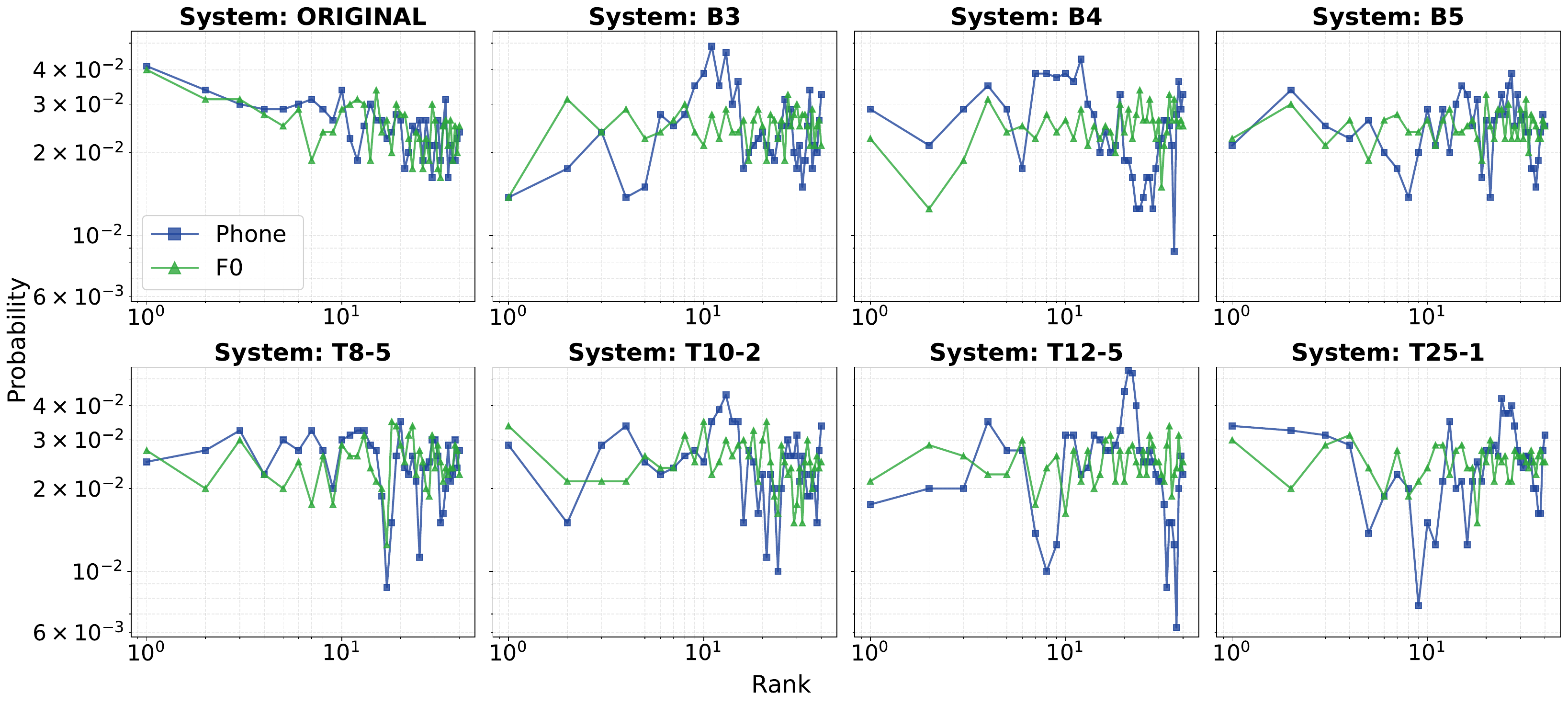}
  \caption{
  As for Figure~\ref{fig:ET_W-NT_org} except for fundamental frequency and phone embeddings.
  }
  \label{fig:f0_phoneme_hist}
\end{figure*}

\subsubsection{Non-timbral embeddings (W-NT)}

Recent work~\cite{gengembre2024disentanglingprosodytimbre,bakari2025influencenontimbralcues} shows that non-timbral cues such as prosody, rhythm, speaking style, and accent can still support speaker re-identification after anonymisation.
First, the authors use the self-supervised learning based WavLM model \cite{Chen2022-wavlm} for speech processing. 
Second, to better capture non-timbral cues, the model is fine-tuned using data processed using voice conversion (VC) to manipulate or obfuscate predominantly timbral characteristics.
It is assumed that more non-timbral speaker-dependent cues such as rhythm and speaking rate are preserved and are then captured.
Again, we use two models.  
The first, denoted as W-NT (WavLM, non-timbral), is trained using original data. 
The second, denoted as ${W\text{-}NT}_\text{anon}$, is trained using anonymised data under the same \textit{semi-informed} attack setting.
Both models were provided by the authors of~\cite{bakari2025influencenontimbralcues}. 
The empirical probability distribution is derived in the same way as for ECAPA-TDNN embeddings.

\subsubsection{Fundamental frequency}
The fundamental frequency F0 captures the primary periodicity of voiced speech. 
F0 dynamics correlate with speaker-specific traits such as pitch range, prosodic style, and vocal characteristics. 
We estimate F0 using the pYIN algorithm~\cite{mauch2014pYIN}, an improved version of the original YIN algorithm~\cite{decheveigne2002yinfundamentalfrequency}. 
As in the original work~\cite{backstrom2025privacydisclosuresimilarity}, we limit the F0 range to between 65 and 450~Hz. 
pYIN estimates pitch lags as integers, which results in 107 distinct fundamental frequency values. We estimate F0 using frames of 30~s with a stride of 10~ms, thereby producing in the order of 3000 estimates over an interval of 30~s, though, after discarding non-voiced frames, this number is lower in practice. 
To obtain the empirical probability distribution we compute the euclidean distance between normalised F0 histograms for each input and reference.

\subsubsection{Phone embeddings}

Phone-related features capture speaker-specific articulation patterns and phonetic realisations that may persist after anonymization. 
We use a VQ-VAE-based voice conversion model to extract phone features~\cite{vali2023interpretablelatentspace}. 
The models learn discrete acoustic units (pseudo-phones) in an unsupervised manner, enabling
phone-like segmentation without the need for transcriptions or text normalisation. 
Non-speech intervals are first removed using energy-based voice activity detection (VAD). 

Remaining speech is then partitioned into 20 uniform segments from which phone embeddings in the form of codebook histograms are extracted. To obtain the empirical probability distribution we compute the euclidean distance between phone embeddings for each input and reference.

\section{Results}
\label{section:results}

We present rank histograms for features described in Section~\ref{section: features} and differences in qualitative results derived using a stronger semi-informed attack model. 
We then present quantitative results derived using metrics described in Section~\ref{sec:sdr_metrics}, followed by a comparison to results from the use of statistical approximations described in Section~\ref{sec:beta}.

\subsection{ET and W-NT embeddings}

Rank histograms for original speech and anonymised speech produced using each of the seven different anonymisation systems are shown in Figure~\ref{fig:ET_W-NT_org} for $ET_\text{orig}$ embeddings (green profiles) and ${W\text{-}NT}_\text{orig}$ embeddings (blue profiles).  

Profiles for original speech shown to the top left show mostly similar trends.
The comparatively high probabilities for the first histogram bins (rank-1) and the steep, negative slopes show that speakers can be readily identified using either embeddings; 
the true speaker identity is most often in the rank-1 position. 
The comparatively lower probabilities for other bins show that other reference speakers are rarely confused with the speaker in the input.

All other plots show rank histograms for anonymised speech.
Rank-1 values are universally lower than for original speech for both $ET_\text{orig}$ and ${W\text{-}NT}_\text{orig}$ embeddings, indicating that all systems provide some level of privacy.
For systems B5, T12-5 and T25-1, the near-to-flat slopes indicate that other speakers are now more likely to be confused with the original speaker, indicating stronger privacy.
Steeper slopes for system T10-2 and, to a lesser extent, also T8-5, indicate weaker privacy. 
Even if rank-1 probabilities are lower than those for original speech, the negative slopes for each profile indicate that, in most cases, the original speaker remains frequently identifiable. 

Interestingly, while rank-1 values for original speech are slightly higher for $ET_\text{orig}$ embeddings, for all anonymisation systems other than T25-1, they are higher for ${W\text{-}NT}_\text{orig}$ embeddings. 
This result is expected since most anonymisation systems obfuscate predominantly timbral rather than non-timbral cues.
Consequently, 
after anonymisation, the latter rather than the former are often more informative.
That we observe the opposite for the stronger T25-1 system is an indication that it is more successful at obfuscating both timbral and non-timbral cues.

\subsection{Fundamental frequency and Phone embeddings}

Rank histograms shown in Figure~\ref{fig:f0_phoneme_hist} show similar profiles for F0 (green profiles) and phone embeddings (blue profiles). Profiles for original speech shown in the top left plot show trends similar to those for speaker embeddings.  
While rank-1 probabilities are still highest, the negative slopes are less pronounced.
These observations indicate that F0 and phone embeddings are less informative than speaker embeddings.
Profiles for anonymised speech show that all anonymisation systems are successful in increasing the confusion between input and reference speakers and that, in most cases, the original speaker can no longer be identified.

\begin{figure*}[!t]
  \centering
  \includegraphics[width=0.9\textwidth]{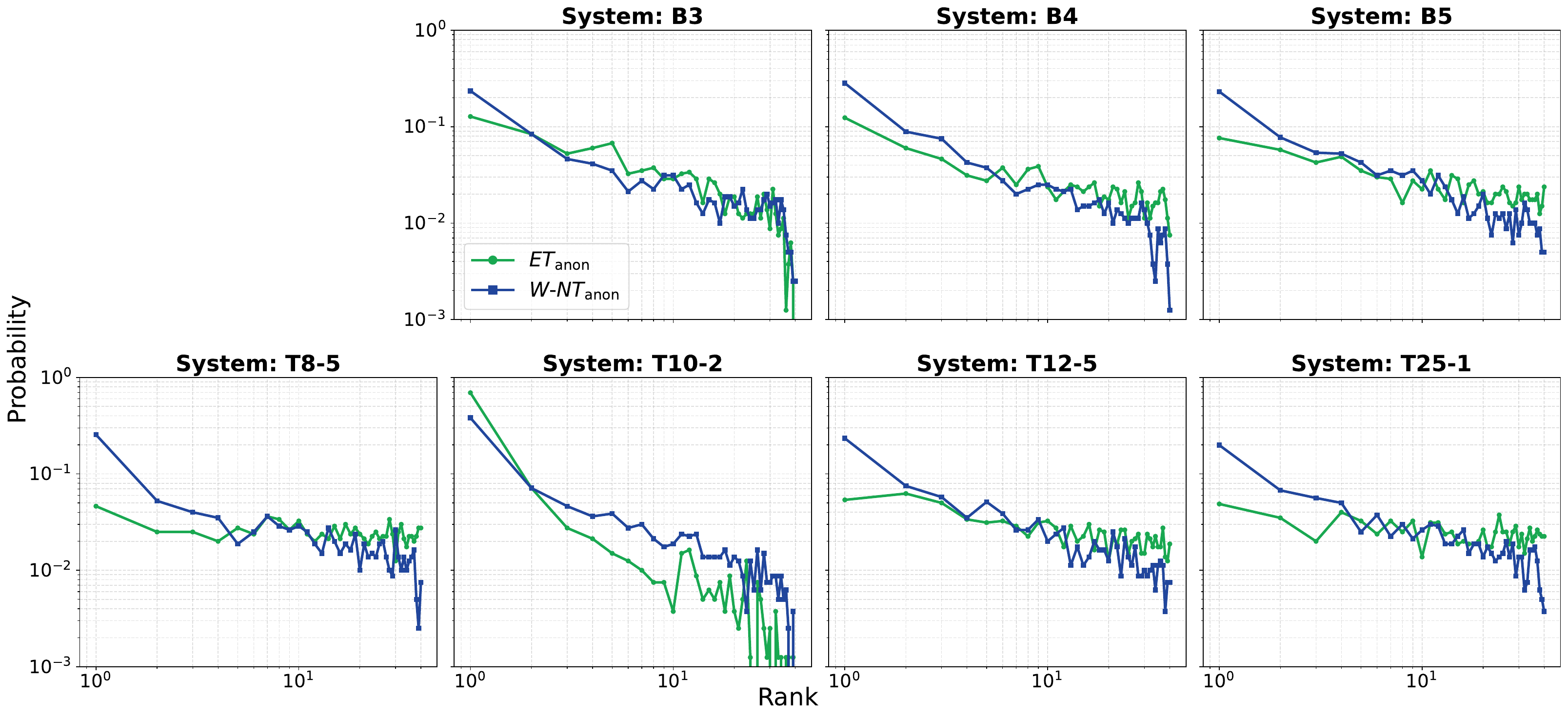}
  \caption{
  As for Figure~\ref{fig:ET_W-NT_org} except for $ET_\text{anon}$ and ${W\text{-}NT}_\text{anon}$ embeddings and the semi-informed attack scenario for which embedding extractors are trained using similarly anonymised speech data.}
  \label{fig:ET_W_NT_anon}
\end{figure*}
\begin{table*}[!t]
  \caption{
  SRD results in terms of maximum (MaxD) and mean (MeanD) disclosure, identification rate (IdR) and rank spread  (RS) for $ET_\text{orig}$ and $ET_\text{anon}$ embeddings.  
  Last column shows corresponding EERs for the 2024 VPC.}
  \label{tab:srd_eer_stats}
  \centering
  \footnotesize
  \setlength{\aboverulesep}{0pt}
  \setlength{\belowrulesep}{0pt}
\begin{tabular}{l|cccc|cccc|c}
    \toprule
    \textbf{System} & \multicolumn{4}{c|}{\textbf{$ET_{orig}$}} & \multicolumn{5}{c}{\textbf{$ET_{anon}$}} \\
    \cmidrule(lr){2-5} \cmidrule(lr){6-10}
    & \textbf{MaxD $\downarrow$} & \textbf{MeanD $\downarrow$} & \textbf{IdR (\%) $\downarrow$} & \textbf{RS $\uparrow$} & \textbf{MaxD $\downarrow$} & \textbf{MeanD $\downarrow$} & \textbf{IdR (\%) $\downarrow$} & \textbf{RS $\uparrow$} & \textbf{EER (\%) $\uparrow$} \\ 
    \midrule
    \textbf{Original} & 4.50 & 2.19 & 56.63 & 7.50 & -- & -- & -- & -- & \textbf{4.6} \\
    \midrule
    B3              & 0.63 & 0.06 & 3.63  & 45.0 & 2.35 & 0.52 & 12.75 & 37.5 & 27.3 \\
    B4              & \textbf{0.53} & \textbf{0.02} & 3.12  & \textbf{47.5} & 2.30 & 0.26 & 12.37 & 25.0 & 30.3 \\
    B5              & \textbf{0.53} & \textbf{0.02} & \textbf{2.50}  & 42.5 & 1.60 & 0.14 & 7.63 & 30.0 & 34.3 \\
    T8-5            & 1.32 & 0.06 & 6.25  & 40.0 & \textbf{0.88} & \textbf{0.03} & \textbf{4.62} & 32.5 & \textbf{40.9} \\
    T10-2           & 3.05 & 0.52 & 20.75 & 20.0 & 4.79 & 3.12 & 69.37 & 7.50 & 40.8 \\
    T12-5           & \textbf{0.53} & \textbf{0.02} & 3.00  & \textbf{47.5} & 1.32 & 0.11 & 5.37 & \textbf{40.0} & 33.2 \\
    T25-1           & 0.67 & 0.05 & 4.00  & 40.0 & 0.96 & 0.05 & 4.87 & 32.5 & 39.8 \\
   \bottomrule
  \end{tabular}
\end{table*}

\subsection{Semi-informed attacks}

All results illustrated in Figure~\ref{fig:ET_W_NT_anon} are computed using embeddings extracted using models trained with anonymised data according to the VPC semi-informed attack model. Rank histograms for embeddings extracted using models retrained with data anonymised by each respective system~\cite{tomashenko2024firstvoiceprivacyattacker,bakari2025influencenontimbralcues} are shown in Figure~\ref{fig:ET_W_NT_anon}.
Profiles are shown for $ET_\text{anon}$ embeddings (green profiles) and ${W\text{-}NT}_\text{anon}$ embeddings (blue profiles). Rank-1 probabilities for $ET_\text{anon}$ and ${W\text{-}NT}_\text{anon}$ models (Fig.~\ref{fig:ET_W_NT_anon}) are higher than those for $ET_\text{orig}$ and ${W\text{-}NT}_\text{orig}$ models (Fig.~\ref{fig:ET_W-NT_org}).
Profile slopes, while near-to-flat for extractors trained using original data, are mostly negative when retrained with anonymised data.
For all but the T10-2 system, rank-1 probabilities are higher for the ${W\text{-}NT}_\text{anon}$ 
model than for $ET_\text{anon}$, confirming recent findings~\cite{bakari2025influencenontimbralcues}.

\subsection{Privacy disclosure}

While the more qualitative findings above provide an insightful picture, rank histograms alone do not offer a means to compare the strength of different systems.  
Quantitative comparisons are made using the SRD metrics described in Section~\ref{sec:sdr_metrics}.
In keeping with VPC policy, and to facilitate performance comprisons, we present results derived using the $ET_\text{orig}$ and $ET_\text{anon}$ models only.\footnote{Results for other models and feature representations are available at \url{obscured\_for\_blind\_review}.} 
Table~\ref{tab:srd_eer_stats} shows maximum disclosure (MaxD), mean disclosure (MeanD), identification rate (IdR) and rank spread (RS) results for both original speech and speech anonymised using each of the seven systems (first column).  
Results for $ET_\text{orig}$ are shown to the left and for $ET_\text{anon}$ to the right. EER results, computed according to the standard VPC protocol under the semi-informed attack model are shown for comparison (last column).

Results again show the importance of using strong attack models for evaluation.  
While, for $ET_\text{orig}$, maximum and mean disclosures are lowest for systems B4, B5 and T12-5, they are lower for the T8-5 system in the case of $ET_\text{anon}$ embeddings.  
The identification rate is lowest for system B5 in the case of $ET_\text{orig}$ embeddings, but for system T8-5 in the case of $ET_\text{anon}$ embeddings.
While systems B4 and T12-5 have the highest rank spread initially, system T12-5 fares best for the strongest attack.
Reassuringly, the better results for T8-5 in most cases correspond to the highest EER, albeit by a small margin.

\subsection{Beta-binomial}

Results for the $ET_\text{anon}$ system derived from approximations to the empirical rank histograms using fitted beta binomial distributions (see Sec.~\ref{sec:beta}) are shown in Table~\ref{tab:srd_metric_spk_cll}.
While there are some numerical discrepancies, trends are identical.  
System T8\mbox{-}5 provides the lowest maximum and mean disclosures and the lowest identification rate, while the rank spread is again lowest for system T12-5, and now also system B3.
It is evident that, despite the low number of speakers compared to the original work~\cite{backstrom2025privacydisclosuresimilarity}, differences between the performance of competing anonymisation systems are substantial enough so that identical trends are derived using either rank histograms directly or statistical, parametric approximations.

\begin{table}[!t]
  \caption{
  As for Table~\ref{tab:srd_eer_stats} except for results derived using beta-binomial approximations to the empirical rank distribution.  $ET_\text{anon}$ only.
  }
  \label{tab:srd_metric_spk_cll}
  \centering
  \footnotesize
  \begin{tabular}{l c c c c}
    \toprule
    \textbf{System} & \textbf{MaxD $\downarrow$} & \textbf{MeanD $\downarrow$} & \textbf{IdR (\%) $\downarrow$} & \textbf{RS $\uparrow$}\\ 
    \midrule
    \textbf{Original} & 4.50 & 2.75 & 56.62 & 15.00 \\
    \midrule
    B3              & 2.35  & 0.59 & 12.74 & \textbf{37.5} \\
    B4              & 2.30  & 0.41 & 12.37 & 35.0 \\
    B5              & 1.60  & 0.12 & 7.62 & 32.5 \\
    T8-5            & \textbf{0.88}  & \textbf{0.02} & \textbf{4.62} & 27.5 \\
    T10-2           & 4.79 & 3.39 & 69.37 & 10.0 \\
    T12-5           & 1.10  & 0.05 & 5.37 & \textbf{37.5} \\
    T25-1           & 0.96  & \textbf{0.02} & 4.87 & 30.00 \\
    \bottomrule
  \end{tabular}
\end{table}

\section{Discussion}
\label{sec:discussion}

The SRD provides revealing insights into the differences in privacy protection for competing anonymisation solutions.  
These go beyond a single snapshot like that provided from estimates of the EER in the form of the mean and worst case disclosure and the rank spread. 
By casting evaluation as an identification problem instead of verification, the SRD allows us to explore not just whether a voice is \emph{sufficiently} dissimilar (to the original voice) after anonymisation, but \emph{by how much} the anonymised voice is confusable with the voices of other speakers.
Comparisons between mean and worst-case disclosure in bits also allows meaningful insights into privacy fairness. 
While we explored speaker embeddings, F0 and phone statistics in this paper, the SRD is readily adapted to representations of accent, of speaker sex, of prosodic features beyond F0, phone durations~\cite{tomashenko25_durationfeatures}, etc., and supports the comparison of privacy disclosure coming from any such source of PII.

In one particular case, the SRD reveals stark differences to EER results:
Table~\ref{tab:srd_eer_stats} shows that, in terms of EER, system T10-2 performs just as well as system T8-5, 
while SRD results uncover a different picture.
Rank histograms for system T10-2 in Figure~\ref{fig:ET_W-NT_org} show a high rank-1 probability, while Table~\ref{tab:srd_eer_stats} shows high maximum and mean disclosures. \textcolor{black}{The identification rate is 70\%. 
The cause of overestimated privacy in case of the EER is now well known~\cite{tomashenko2026thirdVPC} and stems from anonymisation being performed at the speaker rather than utterance level, resulting in a weaker attack.  
Our own, further investigations reveal additional, new insights.
They suspect that enrolment data for the T10-2 system are not anonymised.  
Our assumption is that enrolment utterances are simply re-synthesised with the voice of the original speaker meaning they are effectively unprotected.
A high EER is then the result of mismatched enrolment and trial data rather than strong anonymisation. 
Since we use pooled data (see Sec.~\ref{subsection:protocols}), rather than relying upon distinct enrolment and trial data as for evaluation using ASV and EER estimation, we avoid this potential evaluation pitfall.}

Finally, despite the appeal of the SRD and new insights it provides, we stress the importance of using well-trained, strong attack models.
Without them, just like for the EER~\cite{panariello2025risksdetectionoverestimated}, there is always potential for privacy to be overestimated, no matter what the metric, even the SRD.  
Use of the strongest available attacks hence remains fundamental to reliable evaluation.

\section{Conclusions}
\label{section:conclusion}

We investigated use of the similarity rank disclosure (SRD) for evaluating voice anonymisation, providing an information-theoretic assessment of privacy. 
Compared to the automatic speaker verification equal error rate (EER), the SRD offers a more interpretable and fine-grained characterisation of residual privacy risk.
Results for 2024 VoicePrivacy Challenge systems show that the SRD uncovers privacy leakage and system-specific weaknesses that were not immediately apparent from EER-based evaluation. 
In particular, systems with similar EER performance can exhibit markedly different levels of privacy disclosure, highlighting limitations of current evaluation practices and the need for more comprehensive metrics.
By operating directly on feature representations, the SRD is independent of classifier decisions and can be applied to a wide range of representations that encode personally identifying information. 
The SRD hence offers a flexible and comprehensive tool for the evaluation of voice anonymisation.
\vspace{15pt}
\section{Acknowledgements}

\ifinterspeechfinal
     
     This work was funded by the European Union’s Horizon Europe research and innovation programme grant No 101168193. We would also like to thank Rayane Bakari, Nicolas Gengembre, and Olivier Le Blouch (Orange innovation, France) for providing the pre-trained models for W-NT.
\else
     The Odyssey 2026 organisers
\fi
\bibliographystyle{IEEEtran}
\bibliography{bibliography}

\end{document}